\documentclass[twocolumn]{aastex631}
\usepackage{appendix}
\usepackage{natbib}
\usepackage{graphicx}
\usepackage{booktabs,threeparttable}
\usepackage{tabularx}
\usepackage{enumitem}
\usepackage{hyperref}
\usepackage{mathtools, amsmath,amstext}
\usepackage{cleveref}
\usepackage[T1]{fontenc}
\usepackage[]{mdframed}
\usepackage[figure,figure*]{hypcap}

\submitjournal{AJ}
\accepted{February 2025}

\shorttitle{Reevaluation of the USP Classification Boundary}
\shortauthors{Goyal \& Wang}

\begin{document}

\title{Statistical Reevaluation of the USP Classification Boundary: \\ Smaller Planets Within 1 Day, Larger Period Ratios Below 2 Days}

\author[0000-0001-9652-8384]{Armaan V. Goyal}
\affil{Department of Astronomy, Indiana University, Bloomington, IN 47405}

\author[0000-0002-7846-6981]{Songhu Wang}
\affil{Department of Astronomy, Indiana University, Bloomington, IN 47405}

\correspondingauthor{Armaan V. Goyal}
\email{armgoyal@iu.edu}

\begin{abstract}
\noindent
Terrestrial worlds with $P < 1$ day, known as ultra-short period planets (USPs), comprise a physically distinct population whose origins may be attributed to various possible formation channels within multi-planet systems. However, the conventional 1 day boundary adopted for USPs is an arbitrary prescription, and it has yet to be evaluated whether this specific cutoff, or any alternatives, may emerge from the data with minimal assumptions. We accordingly present a statistical evaluation of the USP classification boundary for 376 multi-planet systems across \textit{Kepler}, \textit{K2}, and TESS. We find that USPs are smaller in size ($p = 0.004$) and exhibit larger period ratios with their immediate neighbors ($\mathcal{P} = P_{2}/P_{1}$; $p < 10^{-4}$) when compared to non-USP short-period ($1 < P/\text{days} < 5$) worlds, and that these discrepancies rapidly transition towards statistical insignificance ($p > 0.05$) at respective orbital periods of $P_{R} = 0.97^{+0.25}_{-0.19}$ days and $P_{\mathcal{P}} = 2.09^{+0.16}_{-0.22}$ days (see Figure \ref{fig3}). We verify that these results are not driven by imprecise planetary parameters, giant companions, low-mass host stars, or detection biases. Our findings provide qualitative support for pathways in which proto-USPs are detached from companions and delivered to $P \lesssim 2$ days via eccentric migration, while a subset of these objects near $P \sim 1$ day experience subsequent orbital decay and refractory mass loss to become USPs. These results lend evidence towards an astrophysical basis for the 1 day USP cutoff and encourage consideration of an additional 2 day boundary within future investigations of USP architectures and evolutionary dynamics.
\end{abstract}

\section{Introduction}\label{intro}
The era of space-based transit photometry heralded by NASA's \textit{Kepler} mission \citep{borucki} and the Transiting Exoplanet Survey Satellite (TESS; \citealt{ricker}) has witnessed the discovery and characterization of thousands of individual extrasolar worlds, among which those between the size of Earth and Neptune have readily emerged as the most common type of planet in the Milky Way (e.g. \citealt{akeson}; \citealt{johnson}; \citealt{stassun}). These prototypical, ``small'' worlds most often occupy multi-planet systems with close-in ($P \lesssim 100$ days), tightly-packed, coplanar orbits (\citealt{xie}; \citealt{thompson}), where planets orbiting the same star tend to exhibit a ``peas-in-a-pod'' uniformity in their size, mass, and orbital spacing (\citealt{millholland}; \citealt{weiss}; \citealt{wang_2017}; \citealt{goyal2022}). While the existence and characterization of such a preeminent configuration may allow for powerful constraints to be levied on the most general mechanisms that govern planet formation (e.g. \citealt{adams_energy}; \citealt{goldberg}; \citealt{xu_bump}), similarly valuable insight may be gleaned from atypical systems and worlds that probe the more extreme dynamical pathways achievable for multi-planet systems. 

Perhaps the most well-known example of an atypical subclass within the small-planet population is that of ultra-short period planets (USPs), which are conventionally defined as as worlds with orbital periods $P < 1$ day (\citealt{sahu}; \citealt{sanchis_ojeda}; \citealt{winn2018}). USPs only occur around $\lesssim 1\%$ of stars (e.g. \citealt{sanchis_ojeda}; \citealt{petrovich}), are generally smaller than other subgiant worlds ($R_{p} \lesssim 2 R_{\oplus}$; e.g. \citealt{rappaport}), and are typically observed to be significantly detached from rest of their associated planetary system (\citealt{steffen2013}; \citealt{steffen2016}). While the rarity, orbital characteristics, and isolation of USPs are similar to the respective qualities of hot Jupiters (e.g. \citealt{dawson}), USPs tend to heavily favor residence in systems with multiple planets (\citealt{adams}; \citealt{qian}), and around hosts with solar or sub-solar metallicity (\citealt{winn2017}), indicating that they likely harbor no causal relationship with hot gas giants and are not the evaporated cores of these worlds. Nevertheless, a migratory history is heavily favored for such worlds, as current USP orbital periods lie within the dust sublimation radius (e.g. \citealt{millholland2020}) while the bulk densities of these planets are consistent with the Earth-like or iron-rich planetary compositions of terrestrial worlds on more distant orbits (\citealt{rappaport}; \citealt{dai2019}; \citealt{uzsoy}). 

While it is thus clear that USPs comprise a population that is architecturally and dynamically distinct from the general small-planet regime, further understanding of their possible origins is precluded by the fact that the conventional 1 day boundary for USP classification is entirely arbitrary, and has largely been adopted due to its simplicity rather than its correspondence to any known underlying astrophysical transition (\citealt{sahu}; \citealt{sanchis_ojeda}; \citealt{winn2018}). As such, it remains to be directly verified if this 1 day boundary is truly the most appropriate means of characterizing the USP population, or if any alternative cutoffs may emerge from evaluation of the data with minimal prior assumptions. 

Motivated by this long-standing uncertainty, we perform in this work a blind statistical search for cutoff orbital periods above which the physical and architectural signatures of the USP population are no longer present. Utilizing a sample of 376 multi-planet systems from \textit{Kepler}, \textit{K2}, and TESS with an innermost period $P < 5$ days, we find that, compared to a control group of worlds with $3 \leq P/\text{days} < 5$, planets are systematically smaller in size up to $P \approx 1$ day, and remain architecturally detached from the rest of their system until $P \approx 2$ days. We respectively present the construction of our sample and our primary null hypothesis testing framework in Sections \ref{sample} and \ref{stats}, conduct our primary analysis in Section \ref{analysis_results}, provide statistical validation of our results in Section \ref{validation}, and discuss the broader astrophysical implications of our findings in Section \ref{disc}.

\section{Sample Selection}\label{sample}
As described earlier, arguably the two most distinct properties of USPs, beyond their extremely short orbital periods, are their small sizes (e.g. \citealt{rappaport}) and their significant architectural detachment from companion worlds (e.g. \citealt{steffen2013}). Accordingly, planetary radius ($R_{p}$) and the period ratio of the innermost planetary pair ($\mathcal{P} = P_{2}/P_{1}$) may comprise the most robust empirical criteria for our intended purpose of constraining the overall regime of USP-like configurations in a manner agnostic to orbital period. Compared to analogous planetary parameters, measurements of $R_{p}$ and $\mathcal{P}$ are far more suited to use in a population-level analysis due to their significant precision, wide availability, and the relatively assumption-free determination afforded by their intrinsic relationships to the observable quantities associated with transit photometry (e.g. \citealt{borucki}; \citealt{weiss_samp}). In a similar regard, these quantities are also not limited by a priori physical scalings with orbital period (e.g. incident stellar flux), the necessity of additional spectroscopic follow-up observations (e.g. planetary mass), or model-dependent assumptions of interior structure (e.g. planetary composition). 

To construct the largest available sample of short-period planets with companions and precise measurements for both $R_{p}$ and $\mathcal{P}$, we select candidate and confirmed transiting multi-planet systems from \textit{Kepler}, \textit{K2}, and TESS based on following criteria:

\begin{enumerate}
    \item Must contain an innermost planet with $P < 5$ days and at least one additional member within $P < 100$ days
    \item Neither of the innermost two planets may be classified as a false-positive detection 
    \item The innermost two planets must have defined measurements for $P$ and $R_{p} \pm \delta R_{p}$
    \item The innermost planet must lie below the hot Neptune desert ($R_{p} \leq 3 R_{\oplus}$; \citealt{lundkvist}; \citealt{dai2021})
    \item The second innermost planet must be subgiant ($R_{p} \leq 4 R_{\oplus}$) to exclude hot Jupiter or hot Saturn companions (e.g. \citealt{dawson}; \citealt{dong})
\end{enumerate}

We exclusively consider transiting systems not only to promote the aforementioned measurement precision afforded for our parameters of interest, but also to minimize the likelihood of undetected USP companions (\citealt{steffen2016}), and to reduced selection bias favorable towards systems with giant planets (e.g. \citealt{petrovich}), which may themselves harbor unique evolutionary histories (see Section \ref{giant}). We justify our inclusion of candidate objects based on prior demonstrations that \textit{Kepler} candidates in multi-transiting systems harbor an overwhelming likelihood ($>99\%$) of association with a physical planet \citep{lissauer_2012}, though we shall nonetheless provide in Section \ref{candidate} an explicit examination of the influence of these candidates with regard to our primary analysis. Limitation of the innermost world to $P \leq 5$ days promotes overlap with the expected dynamical regime for USP formation pathways (\citealt{steffen2016}; \citealt{petrovich} \citealt{millholland2020}), while simultaneously ensuring that our statistical comparisons are elicited for objects within a relatively narrow range of stellar irradiation (e.g. \citealt{owen_2013}) and detection efficiency (see Section \ref{missing_planet}). The final criterion is imposed to remove systems wherein the orbital evolution of the innermost terrestrial world may have been dominated by a close-in giant, such that the dynamical history of the system as a whole is not representative of typical USP formation pathways (see Sections \ref{giant} and \ref{spacing_disc}). While these configurations are intrinsically rare (e.g. \citealt{dawson}), the architectures presented by confirmed systems such as Kepler-9 \citep{wang_kepler9}, GJ-876 \citep{rivera}, and 55 Cnc e \citep{bourrier} indicate that this population is nonetheless likely sizable enough to warrant explicit exclusion in the manner performed thus. 

For \textit{Kepler}, we limit our consideration to the catalog provided by \citet{lissauer}, which sacrifices the uniformity in data processing pipelines associated with Kepler Data Release 25 (DR25; \citealt{thompson}) to place additional emphasis on the accuracy of individual planetary and stellar parameters, which is carried forth through manual vetting of candidates and the refinement of stellar parameters via the inclusion of parallax measurements from Gaia (\citealt{gaia}; \citealt{berger}) and spectroscopic follow-up data obtained with Keck I (\citealt{petigura}; \citealt{fulton}). We only consider planet candidates that are unanimously not listed as false positives across Data Release 24 \citep{coughlin}, DR25 \citep{thompson}, the DR25 supplement\footnote{\url{https://exoplanetarchive.ipac.caltech.edu/docs/Kepler_stellar_docs.html}}, and \citep{lissauer}. Querying for the criteria enumerated above, we obtain 18 \textit{Kepler} multi-planet systems containing USPs (9 confirmed), and 204 \textit{Kepler} systems (169 confirmed) containing non-USP short-period worlds (hereafter non-USPs; $1 \leq P/\text{days} < 5$).

For \textit{K2}, we conduct our search within the ``\textit{K2} Planets and Candidates'' table\footnote{\url{https://exoplanetarchive.ipac.caltech.edu/docs/data.html} } of the NASA Exoplanet Archive (NEA; \citealt{akeson}). For each system, we adopt all planetary and stellar parameter values from the literature associated with the "default" parameter set listed in the table. If any such parameter values are not represented in the default literature, they are populated from the most recent, non-default literature in the table associated with the system. From our query, we obtain 14 USP systems (12 confirmed) and 63 non-USP systems (46 confirmed) from the \textit{K2} data. 

We compile TESS systems from the ``TESS Candidates'' table of the NEA, obtaining 15 USP systems (7 confirmed) and 60 non-USP systems (16 confirmed). The TESS Candidates Table is largely comprised of values from the revised TESS Input Catalog \citep{stassun}, and therefore is often not reflective of updated parameter estimations for confirmed worlds that have been subject to further observational characterization. As such, for the 23 confirmed TESS systems, we adopt planetary and stellar parameters from the ``Confirmed Planetary Systems Composite Data'' table of the NEA. 

\begin{figure}
\centering
  \includegraphics[width=.45\textwidth]{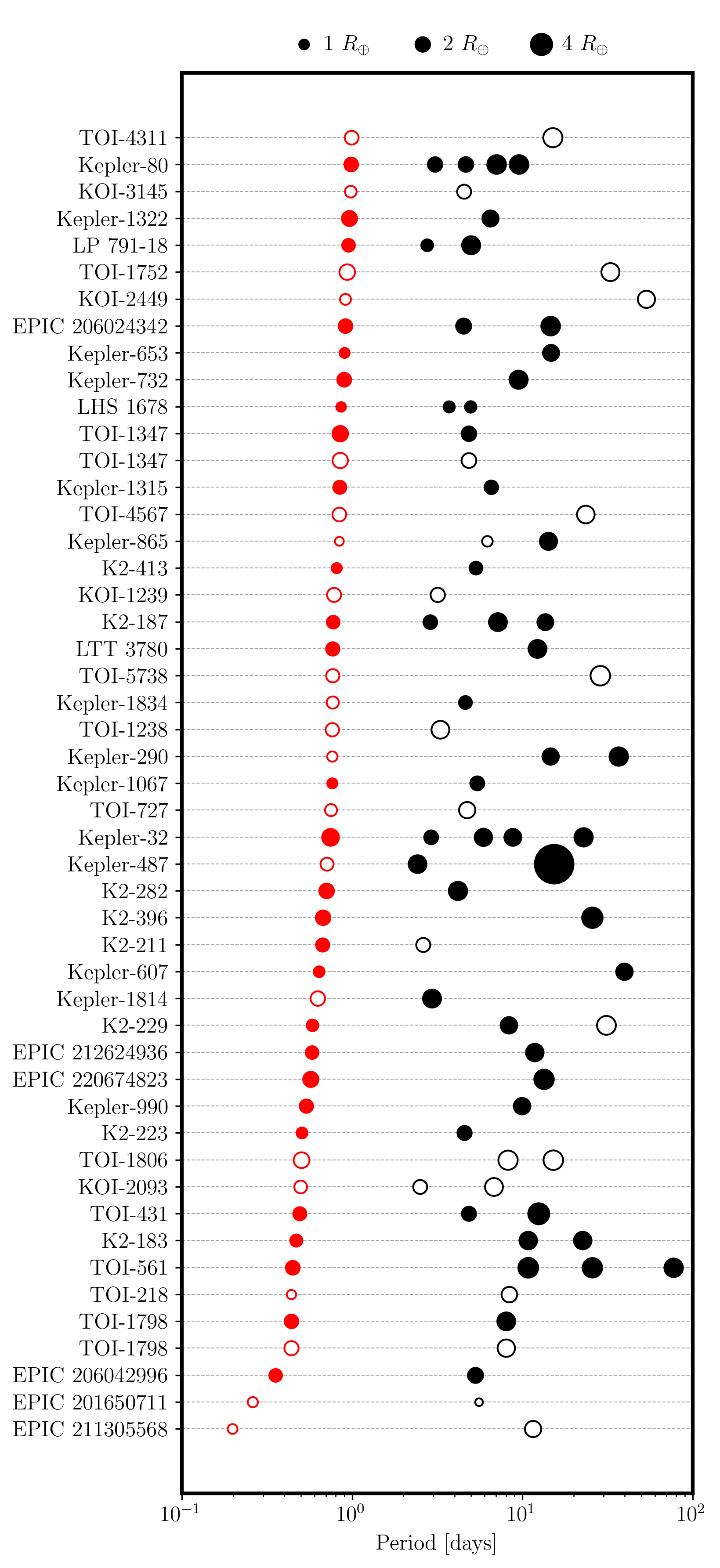}
  \caption{Orbital architectures of the 49 USP systems considered in this work. Marker sizes correspond to planetary radius, filled markers represent objects listed as confirmed within the NEA \citep{akeson}, and hollow markers are associated with \textit{Kepler} candidates from \citet{lissauer}, TESS candidates from from the revised TESS Input Catalog \citep{stassun}, or the NEA's \textit{K2} Planets and Candidates Table \citep{akeson}. We observe, in accord with the literature (e.g. \citealt{rappaport}; \citealt{steffen2013}), that USPs (red) are almost exclusively smaller than $2R_{\oplus}$ and often exhibit wide orbital separations with respect to their non-USP companion planets (black).}
  \label{fig1}
\end{figure}

In total, our sample contains 49 USP systems (26 confirmed) and 327 non-USP systems (205 confirmed). We plot the orbital architectures of the 49 USP systems in Figure \ref{fig1}. We note that as a result of the construction process detailed thus, our overall sample is statistically heterogeneous, and associated population-scale analyses may be subject to complex statistical biases (e.g. \citealt{goyal2022}) resulting from diversity in survey design, instrumental systematics, analysis pipelines, and estimation techniques for individual parameters and their associated uncertainties. While further consideration of these biases and their potential effects lies beyond the scope of this work, we nonetheless wish to preface the forthcoming inquiries with an acknowledgment of such effects. 

Having thus fully treated the construction of our statistical sample, we shall turn now to our primary task of identifying an empirically motivated cutoff period for the USP population. However, given that the basis and possible outcome of such an investigation respectively involve agnosticism and alteration to the established $P < 1$ day definition adopted for USPs, we must proceed with caution with regard to the nomenclature adopted for the various planetary groups compared in our analyses. Accordingly, for the remainder of this work, the term ``USP'' will exclusively refer to objects abiding by the standard $P < 1$ day definition, while the term ``non-USP'' will strictly correspond to the complementary sample of objects with $1 \leq P/\text{days} < 5$. In the considerations of any other cutoffs or separation across a boundary that is not located at $P = 1$ day, we will not make use of the aforementioned labels, and will instead refer to the relevant populations only by an explicit statement of their orbital period regime.

\section{Null Hypothesis Testing Framework} \label{stats}
In order to demarcate the regime where planets no longer harbor the small sizes and architectural detachment of USPs, we must first confirm that the respective size and period ratio distributions of USPs and non-USPs are indeed discrepant in a statistically significant manner, a baseline that has largely been precluded in previous works by the small available sample size of multi-planet systems containing USPs (e.g. \citealt{adams}; \citealt{dai2021}). 

\begin{figure*}
\centering
  \includegraphics[width=.8\textwidth]{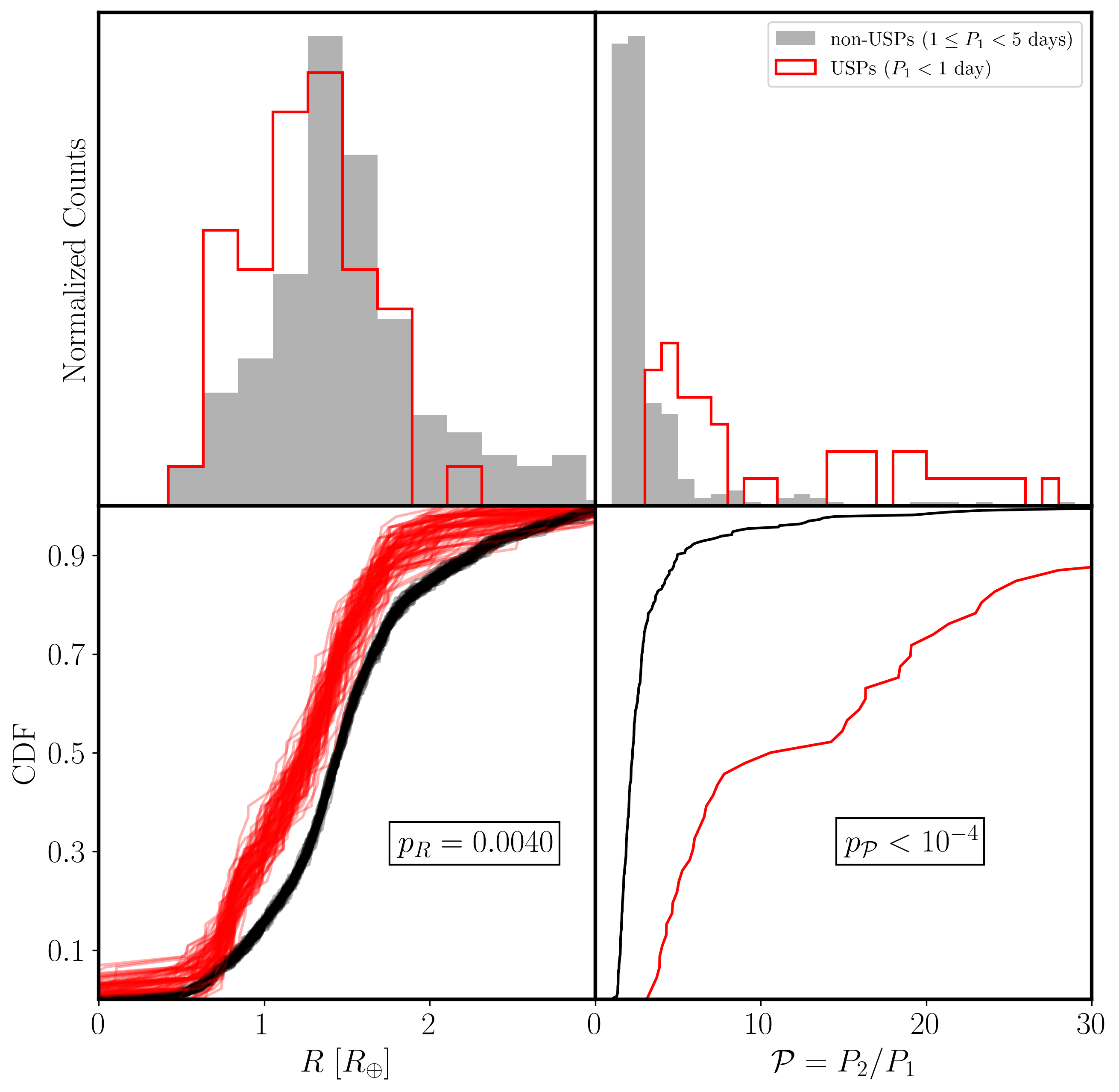}
  \caption{Compared to non-USPs (black/gray), USPs (red) are smaller in size and more architecturally detached from their neighbors in a statistically significant manner. Left: The distribution of 49 USP radii exhibits a relative overdensity at very small sizes ($\sim 1 R_{\oplus}$) and lacks extension beyond $R_{p} \approx 2 R_{\oplus}$ in a manner inconsistent with the 327 non-USP radii and likely representative of refractory mass loss (see Section \ref{size_disc}). We generate 1000 CDFs each for the USP and non-USP population where all individual planetary radii are uniformly resampled within their associated $1\sigma$ uncertainties. Subjecting each of these bootstrapped distributions to a permutation-based implementation of the AD test (see Section \ref{stats}), we find across the 1000 bootstrapping trials a median probability of $p_{R} = 0.0040$ that the USP and non-USP samples are drawn from the same underlying size distribution. Right: USPs are often architecturally isolated from their companions while non-USPs almost exclusively reside in compact systems ($P_{i+1}/P_{i} < 6$; e.g. \citealt{wang_2022}). From the AD test, we obtain a probability $p_{\mathcal{P}} < 10^{-4}$ that the two distributions correspond to the same population.}
  \label{fig2}
\end{figure*}

Perhaps the most conventional means of quantifying the degree of discrepancy emergent for two given distributions is to compute the maximum difference between their cumulative distribution functions (hereafter CDFs) using the two-sample Kolmogorov-Smirnov test (KS test; \citealt{kolmogorov}; \citealt{smirnov}), which has itself risen to fairly ubiquitous usage in various astronomical contexts (e.g. \citealt{hou}; \citealt{feigelson_book}). However, since the endpoint values of a CDF will be 0 and 1 by construction and regardless of the underlying distribution, the KS test has low statistical power with regard to distributions that differ predominantly in their tails (e.g. \citealt{engmann}; \citealt{razali}), which is expected for the size and period ratio distribution pairs for USPs and non-USPs (confirmed in Figure \ref{fig2}). 

As an alternative to the KS test, the two-sample Anderson-Darling test (AD test; \citealt{darling}; \citealt{petit}) performs a similar rank-based calculation, but explicitly places greater weight in the tails of the CDFs. This heuristic difference not only affords the AD test enhanced comparative sensitivity in these particular regions of two given distributions, but also allows for greater statistical power than the KS test even for distributions that have similar tails but differ in shift, scale, or symmetry \citep{engmann}. The AD test may therefore be considered a more robust framework overall with regard to the general comparison of empirical distributions \citep{hou}.

Nonetheless, the calculation of AD test statistic $A_{ij}^{2}$ is complex for $N \lesssim 50$, and determination of associated $p$-values by the AD test, even in modern computational contexts, is often reliant upon approximation or interpolation from tabulated values (\citealt{scholz}; \citealt{feigelson_book}). To mitigate the propagation of numerical errors from these approximations, and to ensure precision in estimation of $p$-values, we utilize the following permutation-based formalism \footnote{Recommended by documentation for SciPy's implementation of AD test: \url{https://docs.scipy.org/doc/scipy/reference/generated/scipy.stats.anderson_ksamp.html}}: For two samples $X_{i}$ and $Y_{j}$ assumed respectively to have continuous parent distributions, we calculate $A_{ij}^{2}$ as the true AD statistic. We then compare $A_{ij}^{2}$ to a distribution of $N_{rand}$ values of $\widetilde{A_{ij}^{2}}$ calculated from randomly drawn mock samples $\widetilde{X_{i}} \in S$ and $\widetilde{Y_{j}} \in S$ where $S$ is the pooled sample $S = X_{i}^{\frown}Y_{j}$. The $p$-value of a shared parent distribution for $X_{i}$ and $Y_{j}$ is then calculated directly as $p(A_{ij}^{2} \leq \widetilde{A_{ij}^{2}})$. We ubiquitously adopt $N_{rand} = 10^{4}$ to maintain exact sensitivity of calculated $p$-values down to $p = 10^{-4}$. However, given that the AD test has been shown to inherently exhibit low statistical power for $N < 30$, and to be overly conservative in this regime even for large-scale differences between two distributions (\citealt{scholz}; \citealt{engmann}; \citealt{razali}), we maintain sample sizes of $N \geq 30$ throughout our analyses whenever possible.

We thereby utilize this permutation-based AD procedure to evaluate the null hypotheses that our 49 USPs and 327 non-USPs share respectively common underlying distribution for $R_{p}$ and $\mathcal{P}$. To appropriately account for measurement uncertainties in planetary size ($\delta R_{P}$), we perform 1000 bootstrap iterations wherein all 376 planetary radii are uniformly resampled within their associated distributions $R_{p} \pm \delta R_{p}$ (see Figure \ref{fig2}), and then calculate a $p$-value from the permutation-based AD test at each one of these bootstrap realizations. We take the median $p$-value across these 1000 bootstrapping realizations as our characteristic probability $p_{R}$, and adopt the same bootstrapping procedure for all calculations of $p_{R}$ throughout this work. Since fractional measurement errors for planetary orbital periods are exceptionally low ($\delta P/P \lesssim 10^{-5}$) for modern space-based transit photometry \citep{lissauer}, we disregard uncertainties in $\mathcal{P}$ throughout this work and calculate characteristic $p$-values $p_{\mathcal{P}}$ from a single run of the permutation-based AD test.

For our full USP and non-USP samples, we obtain $p_{R} = 0.0040$ and $p_{\mathcal{P}} < 10^{-4}$  (Figure \ref{fig2}), thereby allowing for a rejection of the null hypotheses and motivating the search for critical orbital periods $P_{R}$ and $P_{\mathcal{P}}$ where these observed discrepancies between USPs and non-USPs cease to be statistically significant.  

\section{Analysis and Results} \label{analysis_results}
In our efforts to empirically constrain the persistence of the primary physical and architectural signatures of USP-like configurations, the sole assumption we shall levy is that these signatures themselves are expected to vanish within $P \lesssim 3$ days, a notion that is thoroughly supported by both observed and simulated system architectures (e.g. \citealt{lee}; \citealt{petrovich}). In this regard, we shall perform our blind search for the associated USP cutoff orbital period(s) across the 199 total systems with $P_{1} < 3$ days, while the 177 remaining systems with $3 \leq P_{1}/\text{days} < 5$ are maintained as a fixed, non-USP control sample against which the degree of discrepancy in $R_{p}$ and $\mathcal{P}$ will be assessed. This bifurcation further prevents the double counting of systems during our analysis. To affirm the utility of this partitioning, we repeat the null hypothesis AD testing from Section \ref{stats} (Figure \ref{fig2}) but compare the USPs ($P_{1} < 1$ days) only to the control non-USP group ($3 \leq P_{1}/\text{days} < 5$) as opposed to the full non-USP sample ($1 \leq P_{1}/\text{days} < 5$). From this procedure, we observe a general invariance in the obtained characteristic probability values ($p_{R} = 0.0025$ and $p_{\mathcal{P}} < 10^{-4}$), thereby validating the construction and implementation of our control sample as proposed.

Our primary analysis will be therefore conducted via application of null hypothesis AD procedure to moving bins of systems with $P_{1} < 3$ days: for the 199 total systems in this regime ($N_{P_{1}<3} = 199$), we sort systems by increasing $P_{1}$ and define bins with edges $[P_{i}, P_{i+N_{bin}}]$ for $i \in [0, N_{P_{1}<3} - N_{bin}]$. Throughout this work, we adopt a fixed bin size as $N_{bin} = 30$ to ensure adequate statistical power of the AD test (see Section \ref{stats}) while maintaining the greatest possible sensitivity to any emergent transitions. Such a prescription also operates in tandem with our fixed control sample to mitigate any statistical biases that may arise from variations in sample size (e.g. \citealt{goyal2022}; \citealt{goyal2023}). We compare the $R_{p}$ and $\mathcal{P}$ distributions for the 30 systems each bin to those of the 177 systems in non-USP control sample. As described in Section \ref{stats}, we maintain calculation of $p_{R}$ from 1000 bootstrap iterations of the permutation-based AD null hypothesis procedure, while  $p_{\mathcal{P}}$ is determined from a single iteration of same scheme.

We define critical orbital periods $P_{R}$ and $P_{\mathcal{P}}$ as the last bin centers where $p_{R}$ and $p_{\mathcal{P}}$ respectively cross above $p = 0.05$, a regime where we consider the bin distributions as failing to reject the null hypothesis of being indistinguishable from the non-USP distributions. 
It is essential to acknowledge that the conventional $p < 0.05$ prescription for statistical significance has been shown to promote overestimation of the degree of discrepancy present between two distributions, and may thereby lead to the promulgation of irreproducible population-level conclusions (\citealt{feigelson_book}). Accordingly, the statistical community at large has recently advocated for more stringent prescriptions of $p < 0.01$ or $p < 0.005$ (\citealt{wasserstein2016}; \citeyear{wasserstein2019}) to ensure robust comparison between samples. 

In light of these caveats, we justify our usage of a $p = 0.05$ threshold by affirming that, from a conceptual standpoint, the large-scale binning analysis presented here effectively operates in reverse when compared to a single iteration of the AD test. With regard to the respective size and spacing distributions of USPs and non-USPs, we have already provided evidence (Figure \ref{fig2}) against the null hypothesis of shared origin that is wholly consistent even with the most stringent propositions ($p < 0.005$) of \citet{wasserstein2016} and \citet{wasserstein2019}. As such, our primary binning analysis is concerned not with quantifying the discrepancies between these two populations, but identifying when these discrepancies are themselves no longer statistically significant. To this end, where $p < 0.05$ may act as too liberal a regime for evaluating difference between two given distributions, $p > 0.05$ becomes a conservative threshold for their convergence.

\begin{figure*}
\centering
  \includegraphics[width=\textwidth]{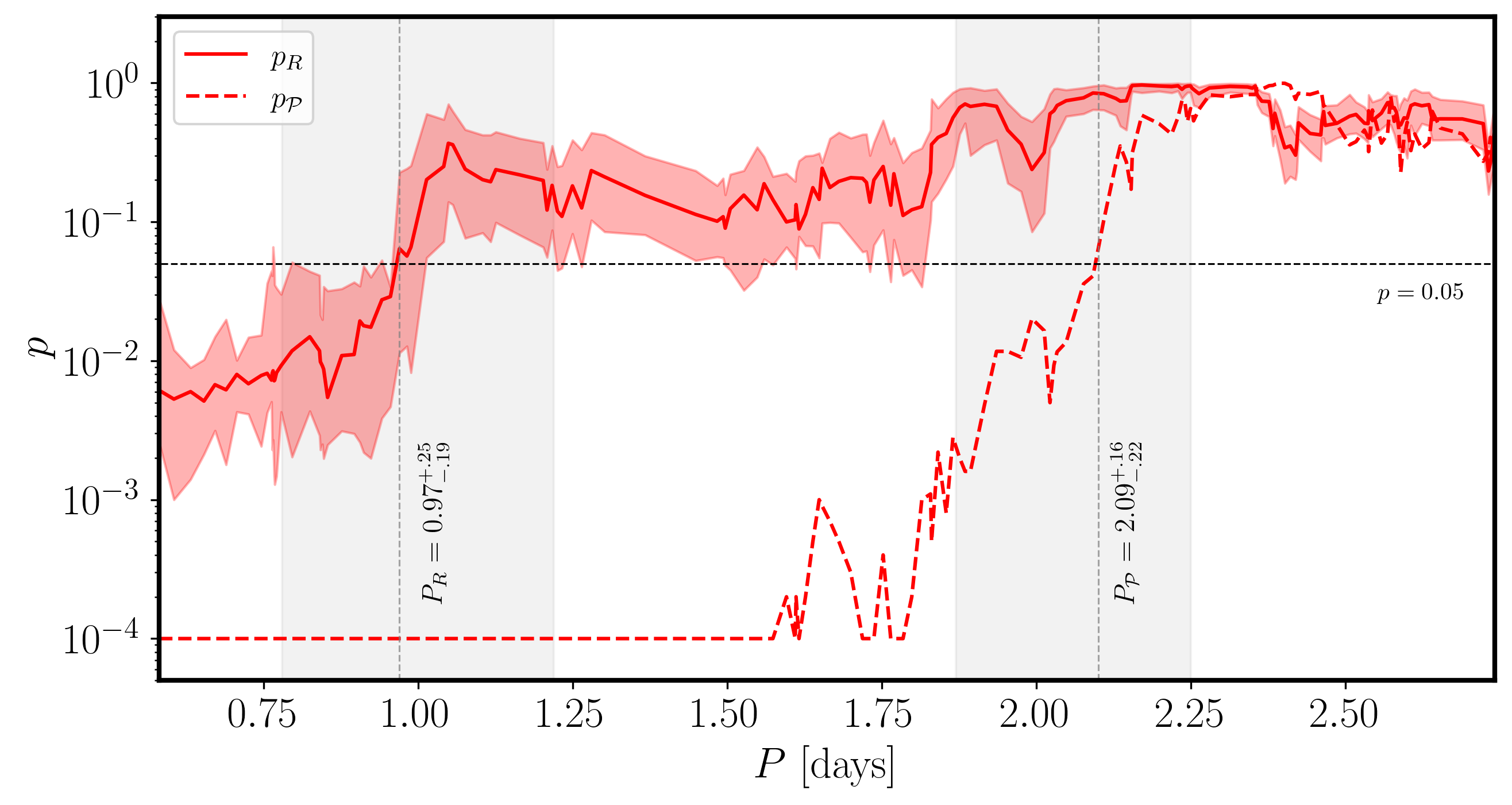}
  \caption{Planets are systematically smaller in size for $P \lesssim 1$ day and more detached from their companions for $P \lesssim 2$ days. Pooling our USP and non-USP samples to remain agnostic to the conventional 1 day USP classification boundary, we use
  the permutation-based AD test to assess the level of statistical discrepancy present between the 199 systems with innermost period $P_{1} < 3$, sorted into moving bins of 30 systems each, and a non-USP control sample of the remaining 177 systems with $3 \leq P_{1}/\text{days} < 5$ (see Section \ref{analysis_results}). The discrepancy observed in planetary size (solid red line) is rendered statistically insignificant ($p > 0.05$, above dashed black line) at $P_{R} = 0.97_{-0.19}^{+0.25}$, while the analogous discrepancy in period ratio (dashed red line) disappears beyond $P_{\mathcal{P}} =  2.09_{-0.22}^{+0.16}$. The former value lies in substantial accord with the conventional 1 day USP classification boundary, while the latter substantiates the existence of a distinct proto-USP population with $1 \leq P_{1}/\text{days} < 2$ \citep{schmidt}. The identified cutoff periods, within their 1$\sigma$ uncertainty ranges, respectively correspond to isolated, monotonic increases of $\sim$2-3 orders of magnitude in $p_{R}$ and $p_{\mathcal{P}}$, indicating that the determined values of $P_{R}$ and $P_{\mathcal{P}}$ are largely insensitive to the choice of critical $p$-value and are themselves representative of underlying transitions between astrophysical regimes (the latter point is discussed in Section \ref{disc}).}
  \label{fig3}
\end{figure*}

We carry forth our binning analysis in full and plot the associated results in Figure \ref{fig3}. Adopting the full bin width as the uncertainty range on our identified transitions, we find critical orbital period values of $P_{R} = 0.97_{-0.19}^{+0.25}$ days and $P_{\mathcal{P}} =  2.09_{-0.22}^{+0.16}$ days after which respective distributions of $R_{p}$ and $\mathcal{P}$ cannot be statistically distinguished from the non-USP control population with $3 \leq P_{1}/\text{days} < 5$. 

The morphology of the observed behavior in both $p_{R}$ and $p_{\mathcal{P}}$ acts as a strong indicator for the astrophysical nature of the identified empirical transitions between the USP and non-USP regimes. 
The $p_{R}$ and $p_{\mathcal{P}}$ curves each exhibit first-order global monotonicity such that $p = 0.05$ is only crossed once in either case, and both exhibit a separation of two clearly statistically distinct regimes, each approximately constant in $p$-value, by a single, abrupt transition. More specifically, these transitions themselves correspond to increases of $\sim 2$-3 orders of magnitude in the span of $\Delta P \lesssim 0.3$ days, and are, to first order, fully encapsulated within the quoted uncertainties on $P_{R}$ and $P_{\mathcal{P}}$. As such, we affirm that the identified cutoffs are largely invariant to the choice of critical $p$-value, as their associated values and uncertainties remain nearly identical for any prescription between $p = 0.01$ and $p = 0.1$, and that the transitions presented here most likely trace underlying evolutionary divergences between distinct planetary populations at $P \lesssim 1$ day, $1 \lesssim P /\text{days} \lesssim 2$, and $P \gtrsim 2$ days. We shall substantiate these claims with thorough statistical validation of our results in the following section and subsequent discussion of their possible dynamical origins in Section \ref{disc}.

\section{Statistical Validation} \label{validation}
\subsection{Influence of Candidate Systems and Imprecise Planetary Radii} \label{candidate}
While such choices are imposed in Section \ref{sample} to allow for provision of the largest possible sample of systems to be considered within our analysis, it is possible that the inclusion of candidate systems and the lack of an explicit restriction on maximum measurement uncertainty may precipitate contamination of our analysis by particularly low-fidelity determinations of planetary radii. As such, we must verify that this broad scheme of inclusion does not propagate forward biases that hold a significant influence over our results.

We thus apply our binning analysis procedure from Section \ref{analysis_results} to the 28 USP systems and 231 non-USP systems where the innermost planet is confirmed, obtaining values of $P_{R} = 0.93_{-0.29}^{+0.34}$ days and $P_{\mathcal{P}} =  2.11_{-0.34}^{+0.24}$ that lie in accord with our primary results (Table \ref{tab1}).

The California Kepler Survey (CKS) presents a nearly homogeneous and high-fidelity catalog of planetary parameters for $\sim 2000$ \textit{Kepler} candidates that were each subject to high-precision spectroscopic follow-up observations and refined estimations of stellar parameters via the consideration of Gaia astrometric data (\citealt{johnson}; \citealt{fulton}; \citealt{weiss}). The uncertainty distribution in planetary size for these CKS planets is almost entirely contained within $\delta R_{p}/R_{p} \leq 0.2$, so we apply this restriction to our sample to promote inclusion of only the most well-characterized USP and non-USP objects. For the 40 USP systems and 298 non-USP systems where the innermost planet has $\delta R_{p}/R_{p} \leq 0.2$, we likewise repeat our binning analysis to obtain characteristic orbital period values of $P_{R} = 1.17_{-0.32}^{+0.38}$ days and $P_{\mathcal{P}} =  2.14_{-0.28}^{+0.15}$ (Table \ref{tab1}).

\subsection{Presence of Giant Companion Planets} \label{giant}
It has been demonstrated thoroughly that systems containing multiple small worlds, particularly those hosted by metal-rich stars, are likely to harbor distant gas giant ($R_{p} \gtrsim 4 R_{\oplus}$) companions (\citealt{zhu_2018}; \citealt{bryan_2019}; \citealt{zhu_2024}). However, there is substantial evidence to suggest that the presence of these giants may inexorably alter the evolution of the inner system, as they may excite significant mutual inclinations for close-in objects \citep{lai}, govern the degree of spacing uniformity observed for the present-day architecture (\citealt{he}; \citealt{kong}), or disrupt the growth of terrestrial worlds in a manner possibly analogous to the solar system (e.g. \citealt{gomes}; \citealt{morbidelli_nice}; \citealt{tsiganis}; \citealt{walsh}). 

In this regard, it is possible that systems with giant companions comprise a dynamically distinct population that may harbor some influence on population-level analyses of small-planet configurations. We thereby perform our primary binning analysis for the 48 USP and 316 non-USP systems where all planets fall within the subgiant ($R_{p} \leq 4 R_{\oplus}$) regime, obtaining characteristic values of $P_{R} = 0.97_{-0.19}^{+0.26}$ and $P_{\mathcal{P}} = 2.08_{-0.22}^{+0.17}$ (Table \ref{tab1}). While this generally affirms that the presence of distant giant companions does not substantially affect the results of this work, we caution that some systems within our sample may host undetected gas giants, though the characterization of such objects may be emerge from long-term radial velocity surveys of small-planet hosts \citep{weiss_kgps}.

\subsection{Presence of Low-Mass Host Stars} \label{m_dwarf}
M-dwarfs are known to host an overabundance of small terrestrial worlds in compact multi-planet chains that are often regarded as ``miniature'' versions of the typical configurations surrounding main-sequence FGK stars (\citealt{dressing}; \citealt{mulders}; \citealt{sabotta}). Such low-mass hosts may also boast a greater overall formation efficiency for USPs (e.g. \citealt{winn2018}; \citealt{millholland2020}), as well as USPs that are less detached from their outer companions (\citealt{petrovich}), perhaps as a result of in-situ formation as opposed to a purely migratory history (\citealt{lee}). Accordingly, it is necessary to ascertain if our USP and non-USP samples indeed exhibit systematic differences in their host spectral type, and if any such differences may act as an astrophysical confounding factor within our results.

To the first point, we compare the distributions of $T_{eff}$ for either sample. We exclude 17 non-USP systems for lacking an associated estimate of $T_{eff}$, while an additional 5 USP and 44 non-USP systems that lack uncertainty estimates and are assigned conservative values of $\delta T_{eff} = 0.065$ to match the maximum value of the total sample. We perform the bootstrapped version of the permutation-based AD test (described in Section \ref{stats}) on 49 USP systems and 310 non-USP systems with measured $T_{eff}$ to obtain a characteristic probability of $p_{T} = 0.03$. While this value does not represent as strong of a discrepancy as compared to $p_{R}$ and $p_{\mathcal{P}}$, we observe that the USP sample nonetheless exhibits a clear overabundance of hosts  with $T_{eff} < 4700$ K (17 out of 49; $\sim35\%$) in comparison to the non-USP sample (79 out of 310; $\sim25\%$), which stands in qualitative accordance with the aforementioned findings of bolstered USP occurrence around M-dwarfs.

Accordingly, we restrict our total sample to FGK main-sequence hosts only ($4700$ K $\leq T_{eff} \leq 6500$ K), and repeat our primary binning analysis for the resulting 32 USP and 231 non-USP systems. We obtain values of $P_{R} = 1.05_{-0.28}^{+0.50}$ days and $P_{\mathcal{P}} =  2.08_{-0.35}^{+0.28}$ (Table \ref{tab1}), indicating that our results are maintained even with the exclusion of M-dwarf hosts.

\subsection{Effects of Observational Bias and Undetected Planets} \label{missing_planet}

\subsubsection{Transit Detection Efficiency}
Any population analysis reliant upon inter-system comparisons of period ratios or other measures of planetary orbital spacing are strongly susceptible to biases from undetected planets that may drive artificial population-level trends across the observed architectures (\citealt{he}; \citealt{goyal2024}; \citealt{rice}). Photometric detection heuristics are also inherently biased towards larger planets such that the physical level of size discrepancy between two populations may be less than what is observed as a result of undetected small worlds within both representative samples (e.g. \citealt{millholland_edge}; \citealt{goyal2024}). Therefore, we shall perform here a rigorous evaluation of transit detection sensitivity across our sample to assess whether the discrepancies in $R_{p}$ and $\mathcal{P}$ recovered by this work are not artifacts respectively shaped by the undetected innermost members of non-USP systems and undetected outer neighbors of USPs. 

Given that the systems considered in this work correspond to various surveys, instruments, and pipelines, we cannot employ a global cutoff in signal-to-noise ratio (SNR) or use unbiased estimates of stellar photometric precision (e.g. \citealt{christiansen}) to directly compare the detection efficiency of small worlds around different hosts. We thus instead adopt a simple heuristic framework where, within each individual system, the SNR of a hypothetical undetected planet is compared to the empirical SNR values of the observed worlds. We consider the approximation for the SNR of a transiting planet in the presence of stellar white noise alone \citep{pont}:

\begin{equation} \label{eq1}
    \text{SNR} = \left(\frac{R_{p}}{R_{*}}\right)^{2}\sigma_{w}^{-1}  \sqrt{\frac{t_{B}t_{0}}{t_{c}P}},
\end{equation}

where $\sigma_{w}$, $t_{b}$, $t_{0}$, and $t_{c}$ respectively denote the Gaussian stellar noise amplitude, total temporal baseline of observation, transit duration, and cadence of data collection. Given that $\sigma_{w}$, $t_{b}$, and $t_{c}$ may be considered invariant for observations of a single host with a single instrument, we may compute the ratios $\mathcal{R}_{i}$ of the SNRs of a hypothetical missing planet $(\widetilde{R_{p}}, \widetilde{P}, \widetilde{t_{0}})$ to the SNRs of the observed planets within a single given system $(R_{p,i}, P_{i}, t_{0,i})$:

\begin{multline} \label{eq2}
    \mathcal{R}_{i} \equiv \frac{\widetilde{(\text{SNR})}}{(\text{SNR})_{i}} = \frac{\widetilde{R_{p}}^{2}\widetilde{t_{0}}P_{i}}{R_{p,i}^{2}t_{0,i}\widetilde{P}}, \\ \text{with } \widetilde{t_{0}} = \frac{\widetilde{P}}{\pi} \arcsin \left[ \frac{R_{*} + \widetilde{R_{p}}}{a(\widetilde{P}, M_{*})} \right],
\end{multline}

where $t_{0,i}$ are the observed transit durations of the detected planets, $\widetilde{P}$ is the hypothetical period drawn from the empirical distribution presented by \citet{neil}:

\begin{multline} \label{eq3}
p(P) \sim
    \begin{cases}
        P^{\beta_{1}} & \text{for } P < P_{b}\\
        P_{b}^{\beta_{1} - \beta_{2}}P^{\beta_{2}} & \text{for } P \geq P_{b}
    \end{cases} \\
    \text{with } P_{b} = 7 \text{ days}, \beta_{1} = 0.90,  \beta_{2} = -0.69,
\end{multline}

$\widetilde{t_{0}}$ is the hypothetical transit duration assumed for a perfectly edge-on $(b = 0)$ transit, and $a(\widetilde{P}, M_{*})$ is the semi-major axis of the hypothetical planet calculated from Kepler's laws. We may therefore say that a hypothetical planet is detectable if its SNR is greater than at least one of the observed worlds within the same system (i.e. $\mathcal{R}_{i} \geq 1$ for any $i$). Under such a framework, we may consider the following mock observational scheme to assess the detectability of randomly placed hypothetical worlds across our sample, and how such worlds, if deemed as undetectable, would alter the intrinsic architectures of their associated systems:

\begin{enumerate}
    \item For each system, draw $\widetilde{P}$ from (\ref{eq3}) such that the following conditions are satisfied:
    \begin{enumerate}
        \item The hypothetical planet must lie at a period ratio greater than 1.2 from any detected planet to preserve system-wide dynamical stability (\citealt{pu_wu}; \citealt{weiss})
        \item The hypothetical planet must lie at a period ratio greater than 2 from any detected USP to ensure a non-resonant pairwise configuration (e.g. \citealt{schmidt})
        \item If these conditions cannot be satisfied, assign $\mathcal{R}_{i} = 1$ for all planets in the system.
    \end{enumerate} 
    \item If $\widetilde{P} \geq 1$ day, draw $\widetilde{R_{p}}$ from 865 measured non-USP radii across all 376 systems. If $\widetilde{P} < 1$ day, draw $\widetilde{R_{p}}$ from the 49 measured USP planet radii. 
    \item If any $\mathcal{R}_{i} < 1$, consider the hypothetical planet as undetected and add the object to the system. If the added hypothetical planet is itself a USP ($\widetilde{P} < 1$ day), reclassify the entire system from the non-USP to USP group. 
    \item If all $\mathcal{R}_{i} \geq 1$, consider the hypothetical planet as detectable, and maintain the observed system architecture and classification.
\end{enumerate}

\begin{figure}
\centering
  \includegraphics[width=.45\textwidth]{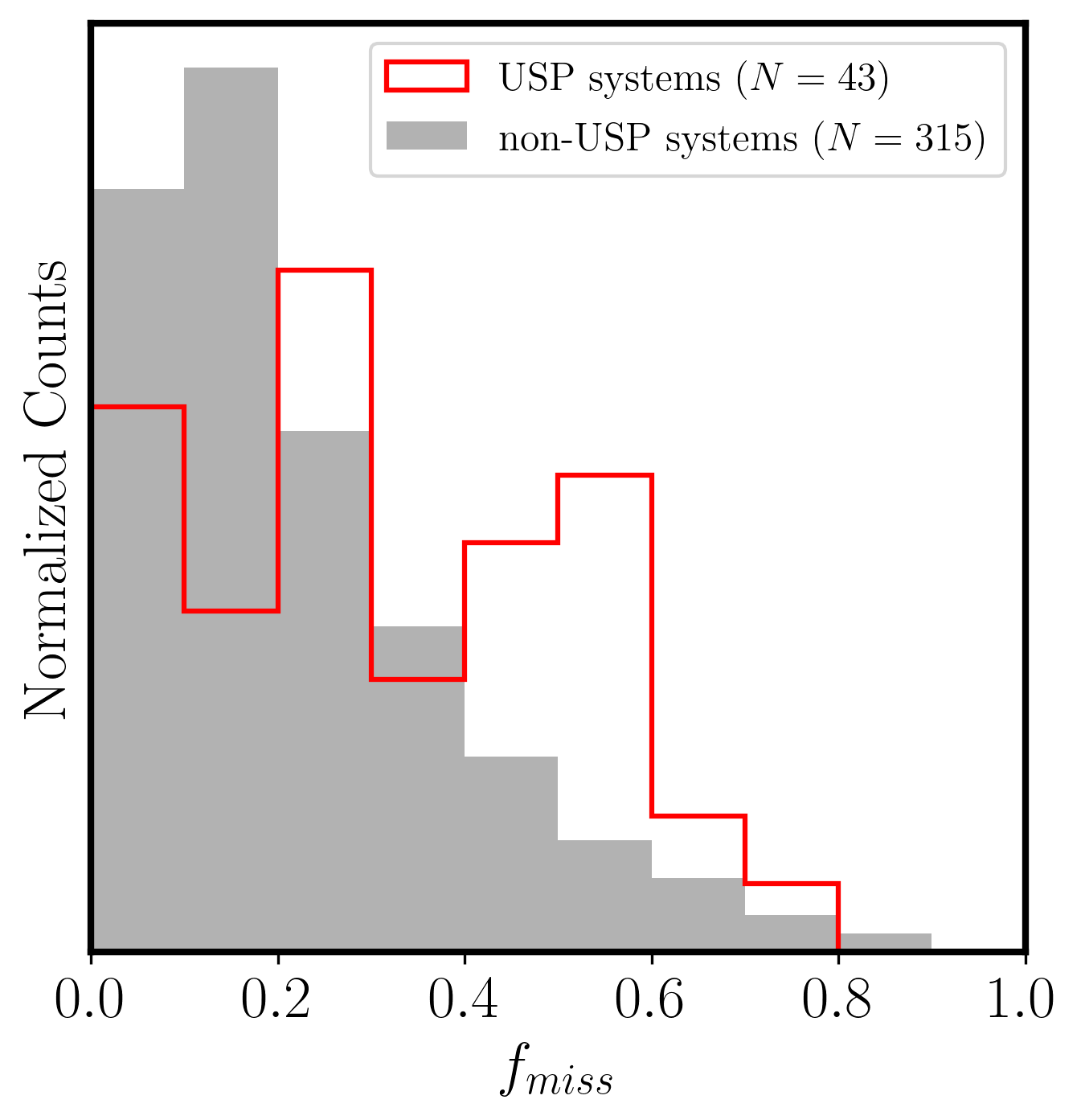}
  \caption{Probability ($f_{miss}$) of an undetected innermost or second innermost planet within each USP and non-USP system. USP systems exhibit a greater likelihood of hosting an undetected companion, in accord with previous findings (e.g. \citealt{steffen2016}; \citealt{qian}), though this likelihood itself falls below $\sim30\%$ for vast majority of both USP and non-USP systems, qualitatively indicating that detection biases are likely of insignificant influence with regard to our primary results. A rigorous empirical investigation of this claim, which includes the calculation of $f_{miss}$, is detailed in Section \ref{missing_planet}: we modify our primary binning null hypothesis procedure (Section \ref{analysis_results} and Figure \ref{fig3}) such that each system is subject to 1000 trials of a mock observational scheme where a randomly generated planet may be added to the inner system architecture if the planet falls below the detection threshold for the system itself. We obtain from this procedure characteristic orbital periods of $P_{R} = 0.92_{-0.15}^{+0.28}$ days and $P_{\mathcal{P}} = 2.16_{-0.19}^{+0.19}$, confirming that our identified transitions in planetary size and spacing are not confounded by missing worlds or detection biases.}
  \label{fig4}
\end{figure}

We note that 158 systems in our total sample lack a stellar mass measurement, and are thus assigned values from the empirical mass-radius relation provided by \citet{demircan} as $M_{*} = (R_{*}/0.89)^{1.12}$. 6 USP and 12 non-USP systems lack measured $t_{0}$ for at least one planet, and are automatically assigned $\mathcal{R}_{i} = 1$ across the system such that their observed system architecture is always maintained.

To simulate the population-scale impact of these detection effects on our identified statistical transitions, we perform a modified version of our primary binning null hypothesis procedure for which this mock observational scheme is applied at each iteration of the bootstrapping process. More explicitly, for each of the 199 bins considered (see Section \ref{analysis_results} and Figure \ref{fig3}), 1000 simulations are conducted where all 914 planetary radii within our sample $R_{p}$ are uniformly redrawn from their associated distributions $R_{p} \pm \delta R_{p}$, the addition of an undetected hypothetical world is then evaluated for all 349 systems via the process outlined above (where systems are reclassified as necessary), and the permutation-based two-sample AD test (see Section \ref{stats}) is finally applied to the modified bin distributions of $R_{p}$ and $\mathcal{P}$.

Carrying forth this modified procedure and evaluating as before the behavior of the median AD-based probabilities in each bin, we obtain characteristic transition values of $P_{R} = 0.92_{-0.15}^{+0.28}$ days and $P_{\mathcal{P}} = 2.16_{-0.19}^{+0.19}$ (Table \ref{tab1}), affirming that our primary results are not substantially influenced by detection biases. For illustrative purposes, we also tabulate for each system the fraction of all bootstrapping iterations in which an undetected planet is added to the system architecture ($f_{miss}$), and observe in Figure \ref{fig4} that USP systems harbor a greater likelihood of undetected transiting companions, in accord with previous findings (e.g. \citealt{steffen2016}; \citealt{qian}), but also that systems in either group typically exhibit a $\lesssim 30\%$ chance of harboring a missing planet.

\begin{table*}
\centering
\caption{Summary of Results$^{1}$ for all Primary and Validative Statistical Tests}
\begin{tabular}{c |c |c||c|c||c| c||c|c||c|c} 
 \multicolumn{5}{c}{} \\
 Description & $N_{USP}$ & $N_{non}$ &  $P_{R}$ & $P_{\mathcal{P}}$ & \multicolumn{2}{c||}{$P_{1} < P_{R}$} &  \multicolumn{2}{c||}{$P_{R} \leq P_{1}< P_{\mathcal{P}}$} &  \multicolumn{2}{c}{$P_{1} \geq P_{\mathcal{P}}$}\\
 \hline
 {} & {} & {} & {} & {} & {$p_{R}$} & {$p_{\mathcal{P}}$} & {$p_{R}$} & {$p_{\mathcal{P}}$} & {$p_{R}$} & {$p_{\mathcal{P}}$} \\
 \hline \hline

Full Sample & 49 & 327 &  $0.97_{-0.19}^{+0.25}$ & $2.09_{-0.22}^{+0.16}$ & 0.0025 & $<10^{-4}$ &  0.057 &  $<10^{-4}$ & 0.49 & 0.36 \\

Innermost Planet Confirmed & 28 & 231 & $0.93_{-0.29}^{+0.34}$ & $2.11_{-0.34}^{+0.24}$ & 0.0035 & $<10^{-4}$ &  0.049 &  $<10^{-4}$ & 0.17 & 0.81\\

Innermost Planet with $\delta R_{p}/R_{p} \leq 0.20$ & 35 & 289 & $1.17_{-0.32}^{+0.38}$ & $2.14_{-0.28}^{+0.15}$ & 0.0006 & $<10^{-4}$ &  0.160 &  $<10^{-4}$ & 0.47 & 0.59\\ 

Exclude Systems with any $R_{p} > 4 R_{\oplus}$ & 48 & 316 &  $0.97_{-0.19}^{+0.26}$& $2.08_{-0.22}^{+0.17}$ & 0.0039 & $<10^{-4}$ &  0.061 &  $<10^{-4}$ & 0.33 & 0.44\\ 

FGK Main-Sequence Hosts Only & 32 & 231 & $1.05_{-0.28}^{+0.50}$ & $2.08_{-0.35}^{+0.28}$ & 0.0031 & $<10^{-4}$ &  0.068 &  $<10^{-4}$ & 0.29 & 0.69\\ 

Include Hypothetical Missing Planets & 49 & 327 & $0.92_{-0.15}^{+0.28}$ & $2.16_{-0.19}^{+0.19}$ & 0.0038 & $<10^{-4}$ &  0.082 &  $<10^{-4}$ & 0.35 & 0.64\\\hline
 
 \end{tabular}
 \begin{tablenotes}
 \small
 \item $^{1}$To prevent double counting of systems (see Section \ref{analysis_results}), all  values of $p_{R}$ and $p_{\mathcal{P}}$ tabulated above are computed with respect to the non-USP control sample defined with $3 \leq P/\text{days} < 5$.
 \end{tablenotes}
 \label{tab1}
\end{table*}

\subsubsection{A Note on Geometric Biases}
\label{inclination}
Given that the entire sample of multi-planet systems treated in our analyses is exclusively constructed from transit observations (Section \ref{sample}), it is possible that even if their constituent worlds are not subject to reduced transit detection efficiency in the manner thus considered, they may still evade characterization as a result of non-coplanar system architectures and the geometric biases emergent therefrom. These concerns are especially pertinent in the context of USP systems since planets with $a/R_{*} < 5$ ($P \lesssim 2$ days for FGKM hosts) have been observed to harbor larger mutual inclinations with respect to their companions ($\gtrsim 6^{\circ}$; \citealt{dai2018}) in comparison to worlds within more prototypical multi-planet systems ($\lesssim 2^{\circ}$; e.g. \citealt{fabrycky}). While a full, statistically rigorous assessment of the impact of such geometric effects on our primary results is non-trivial and lies beyond the scope of this work, we shall nonetheless provide here a brief, qualitative discussion of their expected lack of influence towards our identified population-scale transitions at $P_{R} \approx 1$ day and $P_{\mathcal{P}} \approx 2$ days.

The size transition identified at the former value may be rendered artificial if systems with $1 \lesssim P_{1}/\text{days} < 5$ harbor additional, highly inclined innermost worlds that harbor similarly small radii as objects with $P < 1$ day. While it may be reasoned that some subset of these planets would produce transit signals due to the spatial isotropy with which their individual orbital inclinations would be distributed, it should be noted that all systems treated in this work already contain multiple transiting planets, and thus each exhibit an alignment of their mean orbital plane along our line of sight. Since innermost worlds with $P \sim 1$ day naturally tend to be inclined beyond this mean plane formed by their outer companions (e.g. \citealt{dai2018}; \citealt{becker}), a sample built upon observation of these companion planes is thus inherently and significantly biased against geometries wherein these decoupled innermost worlds would exhibit an edge-on orbit. Nonetheless, given that geometric transit probability increases with proximity to the host star, a smaller companion interior to the observed architectures of these systems may still produce a transit signal for mutual inclinations up to $\sim 10^{\circ}$ (\citealt{dai2018}). Given that this degree of mutual inclination exceeds the typical regime for multi-transiting systems both with and without USPs (e.g. \citealt{fabrycky}; \citealt{dai2018}), it is unlikely that such companions exist, thereby affirming the causal link between the diminutive nature of USPs and their arrival at or persistence within the $P < 1$ day regime (see Section \ref{size_disc}).

In terms of the architectural detachment exhibited by worlds with $P \lesssim 2$ days, we note that regardless of whether this detachment corresponds to a true physical gap in the system architecture or the presence of an additional non-transiting member, the identified statistical boundary at $P \approx 2$ days is still enclosing a population that is morphologically and dynamically distinct from typical compact multi-transiting configurations. This is further corroborated by the notion that the evolutionary pathways most amenable to the detachment of worlds with $P \lesssim 2$ days are also expected to produce large mutual inclinations for these planets (e.g. \citealt{petrovich}; \citealt{pu_lai}), indicating that both properties may act as tracers of the same underlying dynamical mechanisms. We briefly evaluate the role of mutual inclinations in the context of various USP formation pathways in Section \ref{spacing_disc}, and further extend these considerations to single-transiting systems in Section \ref{single}.

\subsection{Confirmation of Large-Scale Statistical Behavior} \label{global}

To ensure that the validative tests performed thus are associated with a similar evolution and scaling of $p$-values as in Figure \ref{fig3} such that their associated values of $P_{R}$ and $P_{\mathcal{P}}$ are not artifacts of a shallow or null transition between the identified regimes, we calculate for each of the supplemental samples considered in this section values of $p_{R}$ and $p_{\mathcal{P}}$ for $P_{1} < P_{R}$, $P_{R} \leq P_{1} < P_{\mathcal{P}}$, and $P_{1} > P_{\mathcal{P}}$. We report this series of values in Table \ref{tab1}, where we observe that all values of $p_{R}$ transition from highly significant levels of discrepancy $(< 0.005)$ for $P_{1} < P_{R}$ to insignificant discrepancy $(\gtrsim 0.05)$ for $P_{R} \leq P_{1} < P_{\mathcal{P}}$, while $p_{\mathcal{P}}$ exhibits a similar transition across $\sim 3$ orders of magnitude from $P_{R} \leq P_{1} < P_{\mathcal{P}}$ to $P_{1} > P_{\mathcal{P}}$. This affirms unanimous consistency of our supplemental investigations with the scaling of $p_{R}$ and $p_{\mathcal{P}}$ displayed in Figure \ref{fig2}, as well as with the global behavior of both parameters in our primary analysis (Figure \ref{fig3}).

We have therefore verified that all values for $P_{R}$ and $P_{\mathcal{P}}$ obtained in this work (Table \ref{tab1}) exhibit a high degree of mutual consistency, and that our primary results of $P_{R} = 0.97_{-0.19}^{+0.25}$ days and $P_{\mathcal{P}} =  2.09_{-0.22}^{+0.16}$ thus do not exhibit significant influence from inclusion of candidate objects, imprecise host parameters, presence of M-dwarf hosts, or biases associated with undetected planets. We have identified an empirical transition in planetary size near $P_{R} \sim 1$ day that displays extremely strong consistency with the traditional USP cutoff in all examined cases, affirming that this boundary likely corresponds to an astrophysical transition in planetary size between the USP and non-USP regimes. We have also presented substantial evidence towards the existence of an analogous transition in the innermost period ratio near $P_{R} \sim 2$ day, which itself may imply the existence of a unique population of worlds with $1 \lesssim P/\text{days} \lesssim 2$ that are not systematically smaller in size but nonetheless exhibit USP-like architectural detachment from their companions. \citet{schmidt} refers to this period range as the ``proto-USP'' regime, so we shall adopt the same nomenclature for the remainder of our work, although we discuss in the proceeding section that such a label does not necessarily imply that objects with $1 \lesssim P/\text{days} \lesssim 2$ will necessarily evolve into USPs. In any regard, we strongly encourage the consideration of both transitions in future inquiries regarding the architectures and dynamical origins of these extremely close-in small-planet systems. If applied jointly, our identified characteristic orbital periods advocate for a classification scheme where:
\begin{itemize}
\item USPs ($P < 1$ day) are both small ($R \lesssim 2 R_{\oplus}$) and exhibit large period ratios ($\mathcal{P} \gtrsim 6$) with respect to their companions,
\item  proto-USPs ($1 \leq P/\text{days} < 2$) may be larger ($R \lesssim 3 R_{\oplus}$) in size but still exhibit comparably large period ratios,
\item Worlds with $P \geq 2$ days exhibit no discernible reduction in planetary size or increase in period ratio.
\end{itemize}
Although there certainty exist individual systems or objects that may challenge these archetypes, such a schematic nonetheless allows for more nuanced considerations than usage of the standard $P < 1$ day cutoff alone.

\section{Discussion} \label{disc}

\subsection{Astrophysical Implications of Size Transition at 1 Day} \label{size_disc}
It is possible that the identified empirical transition in planetary size at $P_{R} = 0.97_{-0.19}^{+0.25}$ days may be sculpted by rapid ($\lesssim 1$ Gyr) atmospheric mass loss associated with XUV-driven photoevaporation or the onset of Roche lobe overflow in such extreme proximity to the stellar host (\citealt{lopez}; \citealt{sanchis_ojeda}; \citealt{koskinen}). However, stellar irradiation remains potent enough to mediate comparably efficient photoevaporation within $P \lesssim 10$ days (e.g. \citealt{owen_2013}), and the initiation of Roche lobe overflow may itself persist up to $P \lesssim 2$ days \citep{koskinen}, both of which indicate that total atmospheric mass loss is not unique to planets within the USP regime. This is further corroborated by the notion that the planetary size distribution for $P < 5$ days (see Figure \ref{fig2}) is overwhelmingly dominated by super-Earths ($1R_{\oplus} \lesssim R_{p} \lesssim 2 R_{\oplus}$), which are not expected to harbor extended atmospheres or significant volatile components (e.g. \citealt{chen_kipping}; \citealt{muller}).

The size transition between the non-USP and USP regimes itself corresponds to a reduced occurrence of super-Earths in favor of sub-Earth ($R_{p} \lesssim 1 R_{\oplus}$) worlds (see Figure \ref{fig2}), and given that both planetary classes are unequivocally associated with barren objects (e.g. \citealt{zeng}), the aforementioned transition may instead by driven by solid mass loss and the evaporation of refractory materials. The dayside temperatures of USPs are generally sufficient to melt rock-forming minerals and most silicates, and it is likely that the surfaces of such worlds harbor substantial magma oceans that may facilitate the release, and possible loss, of silicate vapor evaporated from the planetary crust (\citealt{winn2018}; \citealt{herath}; \citealt{zilinskas}). The proximity of such worlds to the Roche limit of their stellar hosts may trigger tidal disruption events resulting in the infall of refractory material onto the star itself (\citealt{dai_evap}; \citealt{oconnor}), while USPs that have survived tidal disruption for several Gyr may further undergo catastrophic planetary evaporation \citep{curry}. The efficiency of stellar tides also scales dramatically with planetary mass beneath $P \approx 1$ day, such that objects with $M_{p} \gtrsim 2 M_{\oplus}$ may experience total disintegration or engulfment on fairly rapid timescales \citep{lee2025}. It is additionally possible that, in comparison to larger worlds, sub-Earths initially close to $P \approx 1$ day are preferentially drawn into USP orbits by magnetic drag, although such a process would itself nonetheless be accompanied by significant Joule heating and planetary melting \citep{lee2025}. As such, some degree of refractory mass loss is likely an inevitable consequence for objects within the USP regime, though the relative balance of its possible causative mechanisms may be better constrained with further observation of lava worlds, actively disintegrating planets, and the behavior of their host stars (e.g. \citealt{seligman}, see Section \ref{future}).

As a complement or alternative to these mechanisms of mass loss, it may also be the case that the diminutive nature of USPs is a favorable outcome of primordial interplanetary dynamics. If the early evolution of USP systems is primarily mediated by secular interactions (e.g. \citealt{pu_lai}), the resulting equipartition of angular momentum deficit (e.g. \citealt{laskar_2000}; \citealt{laskar_2017}) between planets provides the greatest eccentricity excitation to those objects that are least massive and closest to the star, thereby promoting their efficient inward migration and eventual tidal capture (\citealt{wu_2011}; \citealt{lithwick_2014}). However, such a mechanism would likely imply that USPs should be preferentially smaller than their immediate outer companions, but we do not observe any statistically significant discrepancies between the distributions of $R_{2}/R_{1}$ for USP and non-USP systems ($p > 0.05$ for all bins) that may corroborate these notions.

It should also be noted that particularly small USPs, mainly those in the sub-Earth ($R_{p} \lesssim 1 R_{\oplus}$) regime, may have formed either from especially low-mass protoplanetary disks \citep{xu_bump} or as second-generation objects from a postnebular debris disk \citep{qian}. It thus remains unclear whether some USP-hosting systems were subject to wholly unique dynamical histories as compared to more conventional models of in-situ super-Earth and sub-Neptune formation (e.g. \citealt{terquem}). 

\subsection{Astrophysical Implications of Spacing Transition at 2 Days} \label{spacing_disc}
As opposed to the emergent size transition at $P_{\mathcal{P}} \approx 1$ day, we find that the analogous transition in interplanetary spacing identified at $P_{\mathcal{P}} =  2.09_{-0.22}^{+0.16}$ provides much more stringent constraints on the possible evolutionary pathways for USP systems. We shall thereby assess the qualitative consistency of this 2 day boundary with a variety of formation frameworks, discussed in order of their approximate degree of dynamical quiescence. We shall also make continuous reference to the findings of \citet{schmidt} and \citet{hamer} that USP host ages tend to be greater ($4.7 \lesssim \tau/\text{Gyr} \lesssim 5.8$) than the those of proto-USP worlds ($4.1 \lesssim \tau/\text{Gyr} \lesssim 4.3$) and commensurate with those of typical field stars, as they provide invaluable constraints on the timescales over which USPs may become architecturally detached from their companion worlds.

\subsubsection{Stellar Equilibrium Tides} \label{stellar_tides}
\citet{lee} posits a heavily quiescent model in which nascent planetary chains undergo Type I disk migration up to the magnetospheric truncation radius \citep{terquem}, the innermost planet raises asynchronous equilibrium tides on the host star, and tidal dissipation within the star itself subsequently results in loss of orbital energy and angular momentum for innermost planet. Such a model successfully reproduces observed period distribution of USPs, but may preserve coplanarity too strictly to excite the mutual inclinations observed within USP systems \citep{dai2018}, although this may be alleviated by a subsequent epoch of disruptive postnebular interplanetary interactions  (e.g. \citealt{rusznak}) or nodal precession of the innermost planetary orbit incited by stellar oblateness (e.g. \citealt{li_oblate}; \citealt{becker}). \citet{lee} briefly comment that architectural separation of the innermost planet may indeed persist up to $P_{1} \approx 2$ days, but this remark is strictly qualitative in nature, associated with a small observational sample containing $\lesssim 10$ USP systems, and does not attempt to identify any specific transition between USP and non-USP systems. This empirical trend is also not directly compared to the results of their simulations. Furthermore, \citet{petrovich} demonstrate that, for a typical tidal quality factor $Q'_{*} \sim 10^{7}$ (e.g. \citealt{ogilvie}), the mechanism proposed by \citet{lee} may only promote inward migration within several Gyr for planets with initial periods $P \sim 1$ day, while planets with $P \sim 2$ days may require hosts with unusually strong tidal evolution ($Q'_{*} \sim 10^{5}$). Given that the innermost worlds of newly formed multi-planet systems are expected to halt their migration slightly wide of the magnetospheric truncation radius \citep{masset}, thereby occupying orbits with $P \gtrsim 2$ days (\citealt{mulders_dist}; \citealt{batygin_edge}), stellar equilibrium tides are unlikely to act as the sole progenitor mechanism for USPs, nor can they appreciably incite the level of architectural detachment observed for proto-USPs.

\subsubsection{Planetary Obliquity Tides} \label{obl_tides}
An alternative mechanism proposed by \citet{millholland2020} suggests that, for objects with initial period $P \lesssim 5$ days, tidal dissipation from planetary obliquity tides may force planetary spin vectors into equilibrium ``Cassini states'' with nonzero obliquity, which in turn drives increased tidal dissipation and runaway inward migration in a positive feedback loop. This loop may then be broken once the high obliquity state grows unstable in the presence of sufficiently large dissipative tidal torque from the host star, thus halting inward migration at $P < 1$ day and triggering an abrupt return to a low-obliquity configuration. The associated secular dynamics successfully reproduce the high mutual inclinations observed for USP systems \citet{dai2018}, but the timescale for runaway migration to the final USP orbit is less than 1 Gyr for nominal stellar and planetary masses, and is therefore largely inconsistent with USP ages determined by \citet{schmidt}. Additionally, planets initially near $P \sim 1$ day may require extremely close companions ($\mathcal{P} \leq 1.4$) to trigger sufficient obliquity tides, which lies in tension with the persistence of architectural detachment up to $P \sim 2$ days. Similarly, if such runaway inward migration to USP orbits is highly efficient even for progenitor objects with $P \sim 5$ days, the migration process itself would likely necessitate frequent interruption to generate proto-USPs with their observed level of detachment, the means for which is yet unknown.

\subsubsection{Disk Migration and Weak Eccentricity Excitation} \label{weak_mig}
Reconsidering the primordial migration of resonant chains up to the magnetospheric truncation radius \citep{terquem}, \citet{schlaufman} posits that instead of the stellar equilibrium tides suggested by \citet{lee}, tidal capture of USPs at final orbits with $P < 1$ day may occur as the ultimate result of billions of postnebular secular cycles that gently excite eccentricity ($e \lesssim 0.1$) of the innermost member to promote inward migration. The cumulative timescale for such secular interactions is several Gyr, thereby in accord with observed USP host ages \citep{schmidt}, and while such a process may only yield significant migration if the innermost planet has an additional companion with $P \lesssim 10$ days, we find that this criteria is satisfied by $\sim$70\% of the present-day USP system architectures considered in this work. Such a formation model, like that of \citet{lee}, also achieves consistency with observed distribution of short-period planets given a nominal prescription for stellar tidal efficiency ($Q'_{*} \sim 10^{7}$; \citealt{ogilvie}), though it remains to be confirmed if the observed period ratio distribution (e.g. Figure \ref{fig2}) for this population can itself be reproduced, and if planets with final orbits of $1 \lesssim P/\text{days} \lesssim 2$ may routinely achieve configurations with large period ratios. While the pathways outlined both here and by \citet{lee} may promote coplanarity due to their reliance on quiescent disk migration, the oblateness of their stellar host may induce precession of the innermost planetary orbit in a manner that excites non-trivial inclination without the need for interplanetary scattering (\citealt{li_oblate}; \citealt{becker}). However, the efficiency of such excitations are greatest within the first Gyr of the host lifetime when the star is rotating most rapidly, and it is thus unclear if the associated precession may generate sufficient mutual inclinations for USPs that only arrived within $P < 1$ day after several Gyr of evolution.

\subsubsection{Low-Eccentricity Migration} \label{low_e}
In contrast to the large number of secular cycles necessitated by the prior framework, \citet{pu_lai} put forth a slightly more disruptive ``low-$e$'' ($e \lesssim 0.2$) migration that may be triggered for planets with initial orbital periods from 1-3 days, requiring fewer overall secular interactions with a more massive and eccentric outer companion. While such a mechanism operates over a shorter cumulative timescale ($\sim 1$ Gyr) than that associated with the weaker eccentricity excitations of the \citet{schlaufman} model, the associated secular interactions are expected to occur shortly after the disk dissipation such that subsequent orbital decay (e.g. \citealt{lee}) may occur over several Gyr to achieve final USP orbits at an epoch consistent with observed host ages (\citealt{hamer}; \citealt{schmidt}). This low-$e$ migration further results in a typical degree of USP mutual inclination ($\sim 18^{\circ}$) that is both broadly consistent with the observations of multi-transiting USP systems \citep{dai2018} and the occurrence rates of USPs  that appear as single-planet systems (e.g. \cite{weiss_samp}; see Section \ref{single}). Such a mechanism may also reproduce the observed period ratio distribution for USPs \citep{lee}, as well as the observed trend that average period ratio increases with decreasing period (\citealt{steffen2013}). \citet{pu_lai} remark that their simulated USP and non-USP systems respectively harbor mean innermost period ratios of 14 and 3.5, both of which exhibit correspondence with the values of 15.6 and 2.8 obtained from the sample considered within this work. Furthermore, their simulated proto-USP systems display a mean innermost period ratio between 5.2 and 7, which not only lies in substantial accord with our observed value of 6.2 (see Figure \ref{fig2}), but more broadly suggests, in accord with the findings of this work, that this low-$e$ migration mechanism routinely generates objects with final periods $P \sim 2$ days that are no longer bound in a compact system architecture ($\mathcal{P} < 6$; \citealt{wang_2022}). Such a framework may also result in the production of hot and warm Jupiters for systems containing multiple giant planets \citep{wu_2023}, indicating that low-$e$ migration may serve as a natural mechanism for the delivery of worlds to close-in orbits, regardless of their mass. 

\subsubsection{High-Eccentricity Secular Chaos} \label{high_e}
In a similar, yet more disruptive framework than proposed by \citet{pu_lai}, the model of \citet{petrovich} demonstrates that USP progenitors originally on $\sim 5-10$ day orbits may achieve high-$e$ ($e \geq 0.1$) orbits from a brief epoch of secular chaos before experiencing tidal capture to final orbits within USP regime. Such a pathway necessitates initial high mutual inclinations and predicts final configurations with distant companions, both in a manner consistent with observed USP systems (\citealt{dai2018}; \citealt{adams}). Favoring progenitor objects with initial periods of $P \gtrsim 5$ days, high-excitation secular chaos allows for the most significant inward migration of any mechanism yet discussed, and thereby readily supports  substantial detachment in the final orbital configurations of both USP and proto-USP systems. \citet{petrovich} further remarks that since USPs are only produced in $\lesssim 10\%$ of their simulations, delivery of worlds to proto-USP orbits insufficient for immediate tidal capture is a common outcome of such high-$e$ migration, thus providing a natural explanation for our findings that proto-USPs are architecturally detached from their companions but not systematically smaller in size in the same manner as USPs. In this regard, such a mechanism may operate in tandem with stellar equilibrium tides \citep{lee} such that objects may routinely experience chaotic migration to orbits with $1 \lesssim P/\text{days} \lesssim 2$, while only those nearest to $P \approx 1$ day may incite significant tidal dissipation within the host to experience subsequent orbital decay and refractory mass loss. Given that this high-$e$ migration can be triggered by a relatively small number of secular cycles on a timescale of $\sim100$ Myr, additional orbital decay over several Gyr would imply that USP must be older than proto-USP systems in the manner observed by \citet{schmidt}. However, given that this high-$e$ migration process is only accessible with initial deviations from coplanarity and circularity, and that the associated evolution is itself both chaotic and dynamically disruptive, such a mechanism favors USP configurations with distant ($P \gtrsim 20$ days) companions that harbor large mutual inclinations ($i \gtrsim 20^{\circ}$), which lies in tension with the prevalence of additional close-in ($P \lesssim 10$ days), transiting worlds within USP systems (\citealt{adams}; Figure \ref{fig1} of this work).

\subsection{Single-Planet Systems}
\label{single}
While the empirical transitions identified in this work and their associated astrophysical implications are both robust and generalizable within the context of multi-transiting systems, it is crucial to note that our analyses, by construction, do not consider systems with only a single transiting planet. Such configurations comprise the dominant population within the orbital period regime treated in this work, accounting for $\sim80-90\%$ of observed USPs, and $\sim40-60\%$ of worlds with $1 \leq P/\text{days} < 3$ (\citealt{weiss_samp}; \citealt{pu_lai}). 

This preponderance may be illustrated directly by repeating our sample selection process across \textit{Kepler}, \textit{K2}, and TESS where our first two selection criteria (see Section \ref{sample}) are amended to search only for single-planet systems with $P < 5$ days: the resulting sample contains 154 USP and 901 non-USP systems, the former of which dwarfs our primary sample of 49 USP-containing multi-transiting systems, and both of which are broadly consistent with the aforementioned empirical occurrence rates. However, from application of the permutation-based AD test to the $R_{p}$ distributions of these single-planet systems, we find that single USPs are systematically smaller than single non-USPs ($p < 10^{-4}$), and that their sizes are indistinguishable from those of USPs in multi-transiting systems ($p = 0.645)$, qualitatively indicating that the astrophysical process which govern the observed transition in planetary size at $P \approx 1$ day (see Section \ref{size_disc}) may be invariant to the degree of mutual inclination exhibited by the USP itself.

Analogous consideration of the transition in architectural detachment at $P \approx 2$ days is largely precluded by the inherent reliance thereof on the characterization of at least two worlds within a given system. While it is true, as remarked in Section \ref{inclination}, that the eccentric migration mechanisms most amenable to fostering architectural detachment within $P \lesssim 2$ days may also naturally excite substantially large ($\gtrsim 15^{\circ}$) mutual inclinations for the innermost world (e.g. \citealt{pu_lai}; \citealt{petrovich}, it is also possible that single-transiting configurations may be achieved by more quiescent mechanisms that do not create architectural gaps up to $P \approx 2$ days, such as runaway migration from planetary obliquity tides \citep{millholland2020} or nodal procession from stellar oblateness (\citealt{li_oblate}; \citealt{becker}). Nonetheless, the pathways posited by \citet{millholland2020}, \citet{li_oblate}, and \citet{petrovich}, respectively predict low, moderate, and large stellar obliquities for the innermost planet, thereby indicating that measurement of the spin-orbit angle for planets with $P \lesssim 2$ days may be crucial for constraining their formation histories even in the absence of transiting companions.

It is likewise possible that evolutionary differences between multi-transiting and single-transiting USP systems may be influenced by the existence of the ``\textit{Kepler} dichotomy'', which describes the pronounced overabundance of observed single-transiting systems that lies in tension with classical formation models for multiple-planet systems (\citealt{lissauer2011}; \citealt{hansen}). While this dichotomy may still emerge from a single, continuous distribution of modest mutual inclinations \citep{millholland_tdv}, various works have purported that multi-planet architectures may exhibit an intrinsic bimodality between coplanar configurations and those with large, primordially excited mutual inclinations (e.g. \citealt{ballard}; \citealt{moriarty}; \citealt{zawadzki}). Such a bifurcation may lead to different balances of USP formation pathways in each population, as certain mechanisms may require moderate initial mutual inclinations to transport planets to within $P < 1$ day (e.g. \citealt{petrovich}). Considering these potential divergences between single-transiting and multi-transiting systems, as well as the overall dominance of the former in the context of observed USPs, we shall reaffirm that the results presented in this work ought to be interpreted strictly in the context of the latter population, and we encourage further investigation of both groups under a joint statistical framework.

\subsection{Prospects for Future Work} \label{future}
Across the USP formation pathways considered thus, we find that an evolutionary history composed of a brief epoch ($ \lesssim 1$ Gyr) of appreciably eccentric ($e \gtrsim 0.1$) migration (e.g. \citealt{petrovich}; \citealt{pu_lai}) followed by tidal dissipation within the host star over several Gyr and concurrent refractory mass loss allows for the greatest overall consistency with observed trends in the ages, sizes, and period ratios of both the USP and proto-USP population. Nevertheless, we cannot rule out the long-term, low-$e$ secular evolution posed by \citep{schlaufman}, as it remains to be explicitly demonstrated if this mechanism is qualitatively amenable to the persistence of wide orbital spacing beyond $P \sim 1$ day. 

In any regard, we encourage future dynamical simulations to consider the transition in architectural detachment at $P \approx 2$ days as a crucial constraint for their achieved synthetic configurations. Likewise, simulations that jointly consider the migration, tidal capture, orbital decay, and erosion of USPs may yield significant insight towards the nature of the observed planetary size transition at $P \approx 1$ day. Further observational elucidation is expected via enlargement of the available sample of USPs and proto-USPs in multi-planet systems from the PLantary Transits and Oscillations of Stars mission, which is expected to discover $\sim100$ new worlds with $P \lesssim 2$ days (e.g. \citealt{heller}). 

The vast majority of dynamical mechanisms proposed for the origins of USP and proto-USP worlds assume that the progenitor system consists of first-generation super-Earths formed via convergent migration in the protoplanetary disk (e.g. \citealt{terquem}). However, given that USPs occasionally fall into the sub-Earth ($R_{p} < 1 R_{\oplus}$) regime, it may also be possible, even if remotely, that such objects form instead from especially low-mass disks (e.g. \citealt{veras}) or postnebular debris fields (e.g. \citealt{qian}). To this end, significant insight may be provided by investigations considering a wider variety of initial disk parameters, as well as postnebular evolution that is governed by large-scale mergers and impacts (e.g. \citealt{izidoro}; \citealt{goldberg}; \citealt{lammers}).

The high transit frequency of USPs, as well as their extreme proximity to their hosts, makes them favorable targets for high-precision spectroscopic follow-up with instruments such as the James Webb Space Telescope (JWST) (e.g. \citealt{murgas}; \citealt{seligman}). Given that many USPs are consistent with lava-rich or actively evaporating surfaces, they may act as unique laboratories for assessing the geological evolution, interior structure, and compositional chemistry of terrestrial worlds at large (e.g. \citealt{lee2025}). Follow-up observations of their secondary eclipses with JWST may therefore provide unparalleled insight towards the complex interplay between planetary-scale microphysics and system-level assembly history for both USPs and the broader small-planet population (e.g. \citealt{herath}; \citealt{loftus}). Even beyond direct characterization of the USP itself, the identification of irregular transit signals from planetary dust trails (e.g. \citealt{perez_becker}; \citealt{curry}), host UV activity driven by star-planet magnetic interactions (e.g. \citealt{lanza}; \citealt{shkolnik}), or augmented refractory abundances in the host star (e.g. \citealt{dai_evap}) may each provide further critical insights towards both the planetary and system-scale evolution of such extreme worlds.

\section{Summary and Conclusions}
We have provided in this work a population-level statistical investigation for an astrophysically motivated classification boundary for USPs. Using samples of 49 USP ($P < 1$ day) and 327 non-USP ($1 < P/$days $< 5$) systems from \textit{Kepler}, \textit{K2}, and TESS, we demonstrated that USPs are smaller ($p = 0.004$) and more detached ($p < 10^{-4}$) from their companions compared to non-USPs in a statistically significant manner. 

\begin{mdframed}
    We identify from a blind statistical search across all 376 systems that USPs remain smaller in size up to $P_{R} = 0.97_{+0.25}^{-0.19}$ days and are more detached up to $P_\mathcal{P} = 2.09_{+0.16}^{-0.22}$ days.
\end{mdframed}

We validated that the respective locations of these two transitions are largely insensitive to the precision of our planetary parameters, the inclusion of candidate worlds, the presence of giant companions, stellar host type, and biases induced by possible undetected worlds. Broadly speaking, concurrent consideration of these transitions advocates for a slightly more detailed classification scheme in which USPs are both small and detached from companions, proto-USPs ($1 \leq P/\text{days} < 2$) are not as small but still architecturally detached, and worlds with $P > 2$ days are neither small nor detached.

Our results qualitatively support a USP formation history consisting of initial eccentric migration to a proto-USP orbit , most amenably in the form of the ``low-$e$'' mechanism proposed by \citet{pu_lai}, followed by several Gyr of tidal decay to a final USP orbit \citep{lee} and concurrent refractory mass loss. Such a pathway not only provides a basis for the smaller size of USPs and the dynamical isolation of proto-USPs as identified in this work, but also yields correspondence with prior findings that USP systems likely disfavor coplanar evolution \citep{dai2018}, are typically a few Gyr old \citep{hamer}, and are older than proto-USP systems \citep{schmidt}.

\section*{Acknowledgments} \label{sec:acknowledgments}
The authors would like to thank Brandon Radzom, Jessica Ranshaw, Xian-Yu Wang, Emma Dugan, and Jace Rusznak for their support during the preparation of this manuscript, as well as Kevin Schlaufmann and Sharon Wang for their insightful discussions regarding this work. The authors would also like to extend their fullest appreciation towards the anonymous referee and statistics editor, whose comments and suggestions greatly improved the overall clarity of this work. S.W. gratefully acknowledges the support received from the Heising-Simons Foundation through Grant No. 2023-4050. A.V.G. expresses substantial gratitude towards the financial support received from the Indiana University Bloomington Department of Astronomy through their provision of the Sullivan Graduate Fellowship.

\bibliographystyle{aasjournal}
\bibliography{main}

\end{document}